\documentclass[aps,prmaterials,twocolumn,english,showpacs,superscriptaddress,amssymb,amsfonts,nofootinbib,longbibliography]{revtex4-2}

\usepackage[latin9]{inputenc}
\usepackage{color}
\usepackage{verbatim}
\usepackage{amsmath}
\usepackage{amssymb}
\usepackage{physics}
\usepackage{dsfont}
\usepackage{graphicx}
\usepackage{esint}
\usepackage{esvect}
\usepackage{braket}
\usepackage{amsmath,bm}
\usepackage{graphicx}
\usepackage{subfigure}
\usepackage{url}
\usepackage{lipsum}
\usepackage{lipsum}
\usepackage[dvipsnames]{xcolor}
\usepackage[
    colorlinks,
    linkcolor={blue!80!black},
    citecolor={blue!80!black},
    urlcolor={blue!80!black}
]{hyperref}
\usepackage[resetlabels]{multibib}

\makeatletter
\@ifundefined{textcolor}{}
{%
 \definecolor{BLACK}{gray}{0}
 \definecolor{WHITE}{gray}{1}
 \definecolor{RED}{rgb}{1,0,0}
 \definecolor{GREEN}{rgb}{0,1,0}
 \definecolor{BLUE}{rgb}{0,0,1}
 \definecolor{CYAN}{cmyk}{1,0,0,0}
 \definecolor{MAGENTA}{cmyk}{0,1,0,0}
 \definecolor{YELLOW}{cmyk}{0,0,1,0}
}

\newcommand{\bk}{\mathbf{k}}
\newcommand{\bR}{\mathbf{R}}

\newcommand{\brho}{\boldsymbol{\rho}}
\newcommand{\uvec}{\bm{v}}
\newcommand{\bn}{\mathbf{n}}
\newcommand{\bsigma}{\boldsymbol{\sigma}}

\usepackage{mathrsfs}

\makeatother
\begin{document}

\title{
Symmetry-enforced double Weyl points, multiband quantum geometry, and \\ singular flat bands of doping-induced states at the Fermi level
}
\author{Moritz M. Hirschmann}
\thanks{These authors contributed equally to this work.} 
\affiliation{RIKEN Center for Emergent Matter Science, Wako, Saitama 351-0198, Japan}
\author{Johannes Mitscherling}
\thanks{These authors contributed equally to this work.} 
\affiliation{Department of Physics, University of California,  Berkeley, California 94720, USA}
\date{\today}

\begin{abstract}
    Two common difficulties in the design of topological quantum materials are that the desired features lie too far from the Fermi level and are spread over a too-large energy range.
    Doping-induced states at the Fermi level provide a solution, where nontrivial topological properties are enforced by the doping-reduced symmetry.
    To show this, we consider a regular placement of dopants in a lattice of space group (SG) 176 ($P6\text{}_3/m$), which reduces the symmetry to SG 143 ($P3$).
    Our two- and four-band models feature double Weyl points, Chern bands, Van Hove singularities, nontrivial multiband quantum geometry due to mixed orbital character, and singular flat bands. 
    We relate these features to density-functional theory (DFT) calculations for dopant and vacancy bands of lead apatite Pb$_{10}($PO$_4)_6$O and Pb$_{10}($PO$_4)_6($OH$)_2$, the van der Waals ferromagnet Cr$_2$Ge$_2$Te$_6$, the semiconductor SiC, and the 2D dichalcogenide MoS$_2$.
\end{abstract}

\maketitle

\section{Introduction}
Two difficulties 
in the study of topological band theory \cite{hasan2010, Chiu2016}
are that (i) the topological features \cite{burkov2016, Chan2016, bernevig2018} lie too far from the Fermi level and (ii) 
are spread over a too-large energy range.
Challenge (i) can be addressed by extensive material surveys \cite{vergniory2019, choudhary2019, xu2020}, 
by pushing interstitial electrons into the gap of a molecular crystal \cite{yu2023}, or by defect states, within a gap in the vicinity of the Fermi level \cite{Noh2014, trainer2022, Schultz2006, Kim2023b}.
Difficulty (ii) can be avoided by narrow bands, which are known to occur in hexagonal systems~\cite{nishino2003, Chan2016, liu2020}. 
Systems with narrow and flat bands have attracted broad interest in the last years
\cite{Bistritzer2011, Cao2018, Cao2018a, Phong2023, Sun2011, Han2021, Liu2014, Neves2023, Tian2023, Kang2020, Kim2023, Lau2021, Dutta2023, Liu2022} due to their interplay of quantum geometry and strong correlations.
Besides the well-known Berry curvature, in particular the quantum metric---a distance measure between Bloch states of close-by momenta \cite{Provost1980}---relates to a diverse set of phenomena \cite{Peotta2015,  Holder2020, Mitscherling2020, Komissarov2023, Morimoto2023, Ahn2020, Ahn2022a, Tai2023, Graf2021, Mera2022, Avdoshkin2023, Arbeitman2022, Arbeitsman2022a, Bouzerar2021, Mitscherling2022,Hetenyi2023, Bouhon2023}, which is particularly important in the context of narrow or flat bands \cite{Arbeitman2022, Arbeitsman2022a, Bouzerar2021, Mitscherling2022}. 

To tackle the two aforementioned difficulties in finding geometric narrow bands near the Fermi level, we build on the idea of combining doping-induced in-gap states with symmetry-enforced topology.
Our approach is inspired by density-functional theory (DFT) calculations on pristine copper-doped lead apatite $\text{Pb}_9\text{Cu}(\text{PO}_4)_6(\text{OH})_2$ \cite{Griffin2023} and $\text{Pb}_{9}\text{Cu}(\text{PO}_4)_6\text{O}$ \cite{Kurleto2023, Si2023,Lai2023} that assumed the regular replacement of one of the four lead atoms on the $4f$ Wyckoff positions by copper. 
We show that the DFT band structure \cite{Griffin2023, Kurleto2023} provides precisely an example of the desired scenario with a topological band structure that arises due to the SG 143 ($P3$) of the assumed copper-doped crystal structure.
Whereas this arrangement of dopants has not yet been realized experimentally in copper-doped lead apatite \cite{Puphal2023}, we identify doped Cr$_2$Ge$_2$Te$_6$ \cite{Gong2017, Verzhbitskiy2020, Tang2017}, doped SiC \cite{Kimoto2014, Roccaforte2021, Luo2018}, and sulfur vacancies in MoS$_2$ \cite{Cao2021, Han2019, Loh2021} as candidates of analogous doping-induced topological bands.

To describe the above systems, we construct minimal two- and four-band tight-binding models, for which we identify key topological and quantum geometric features, including double Weyl points, Chern bands, nontrivial Berry curvature, and a nonzero quantum metric for single and combined sets of connected bands due to orbital mixing, and singular flat bands. 
The flatness occurs due to destructive interference \cite{Derzhko2015, Mielke1991, Mielke1991a, Tasaki1998, Bergman2008, Pyykkonen2021, Kim2023, Kobayashi2016, Maimaiti2017, Hausler2015}, which resembles a higher-dimensional version of the Creutz ladder \cite{Creutz1999} and enables further studies of its unique physics  \cite{Bilitewski2018, Bodyfelt2014, Peotta2015, Chan2022, Lin2023, Mao2023b, Hofmann2020, Peri2021, Roy2020, Sayyad2020, Hwang2021, liu2020, Jiang2023, Kitamura2023, Sayyad2023, Muller2016, Mondaini2018, Mahyaeh2022, Mitscherling2022} in combination with the nontrivial multiband quantum geometry.

\section{Minimal tight-binding models}
We consider the rich topology of SG~176 ($P6_3/m$), the SG of lead apatite \cite{bruckner1995, krivovichev2003}. 
It enforces band crossings on points, lines, and planes \cite{bradlyn2017, zhang2018}.
For spinful bands this centrosymmetric group leads to three Dirac nodal lines intersecting at the A point as a result of the off-centered mirror symmetry $m_z$ \cite{yang2017}.
Systems with spinless representations of $P6_3/m$ exhibit always a nodal plane at $k_z = \pi$ \cite{wilde2021} and Dirac points at $k_z =  0$ are possible \cite{bradlyn2017}.

With doping one site per unit cell one should expect that the space group symmetry is in general reduced.
In the absence of a significant structural transition, the new band structure exhibits still the (symmorphic) site symmetry. 
Specifically, substituting one copper on the $4f$ Wyckoff position of lead-apatite, as assumed in DFT \cite{Griffin2023,Si2023,Lai2023,Kurleto2023}, reduces SG~176 to SG~143 ($P3$) with broken inversion symmetry \cite{tokura2018}.
The complex spinless representation of $P3$ provides the minimal model for just a pair of Weyl points at the time-reversal invariant momenta (TRIMs) $\Gamma$ and~A. In contrast, the spinful representations are known to exhibit eight Kramers-Weyl points \cite{chang2018}. 
\begin{figure*}[t!] 
\includegraphics[width = 0.99 \textwidth]{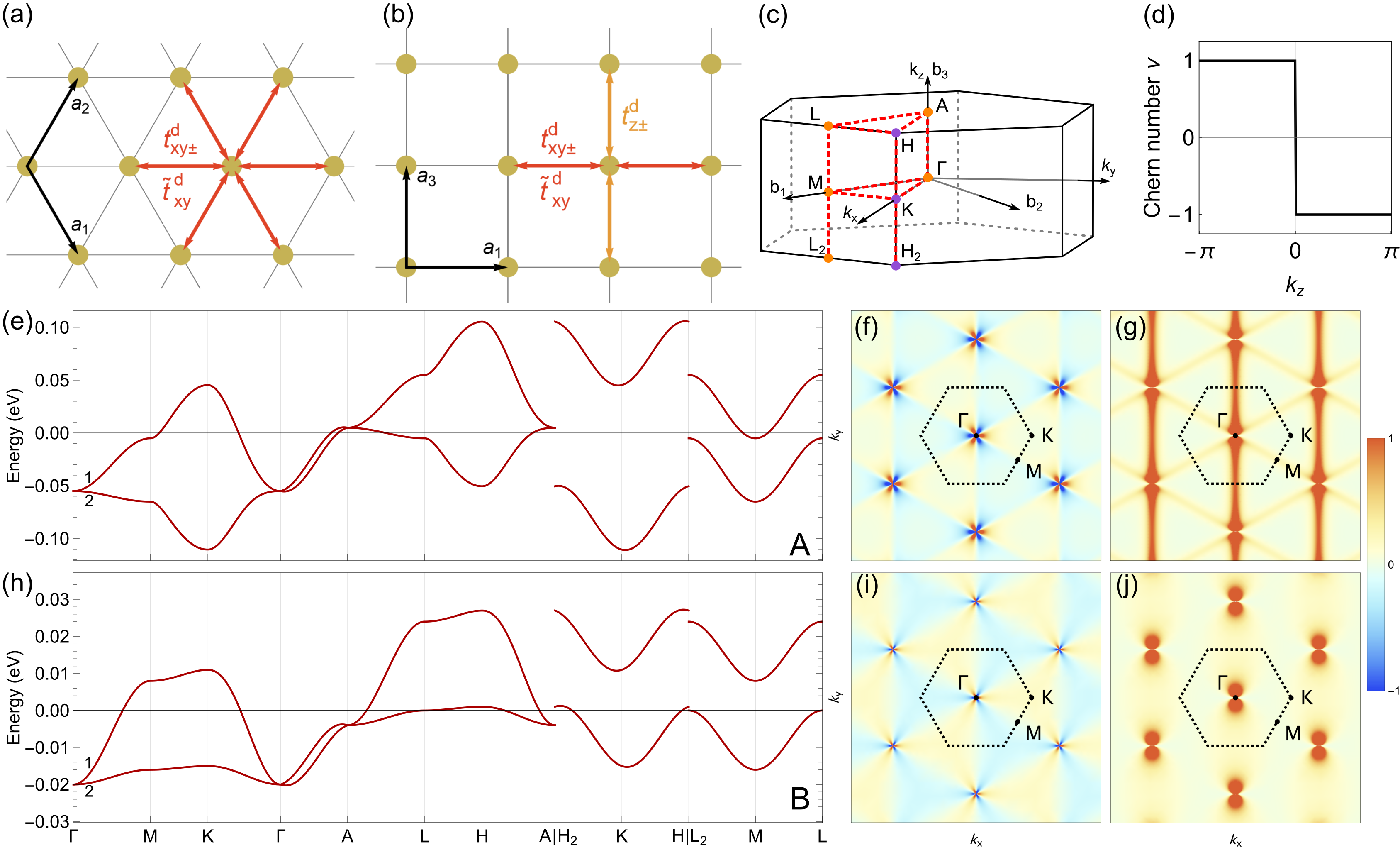}
\caption{\label{fig:TwoBandModel}
\textbf{Two-band model for SG~\textit{P}3}.
[(a),(b)] Top and side view of the trigonal lattice with nearest-neighbor hopping terms, where symmetry-related directions are shown in the same color. 
(c) Brillouin zone with TRIMs (orange), high-symmetry points (purple), and k-path (red) that is used for the band structures with parameter sets A and B in (e) and (h), respectively. The parameters are summarized in Table~\ref{tab:parameters}.
(d) Symmetry-enforced Chern numbers on planes of constant $k_z$.
Berry curvature $\Omega^{xy}_2$ [(f),(i)] and quantum metric $g^{xx}_2$ [(g),(j)] at $k_z = 0$ corresponding to the models to the left.
}
\end{figure*}    

\begin{figure*}[t!] 
\includegraphics[width = 0.99 \textwidth]{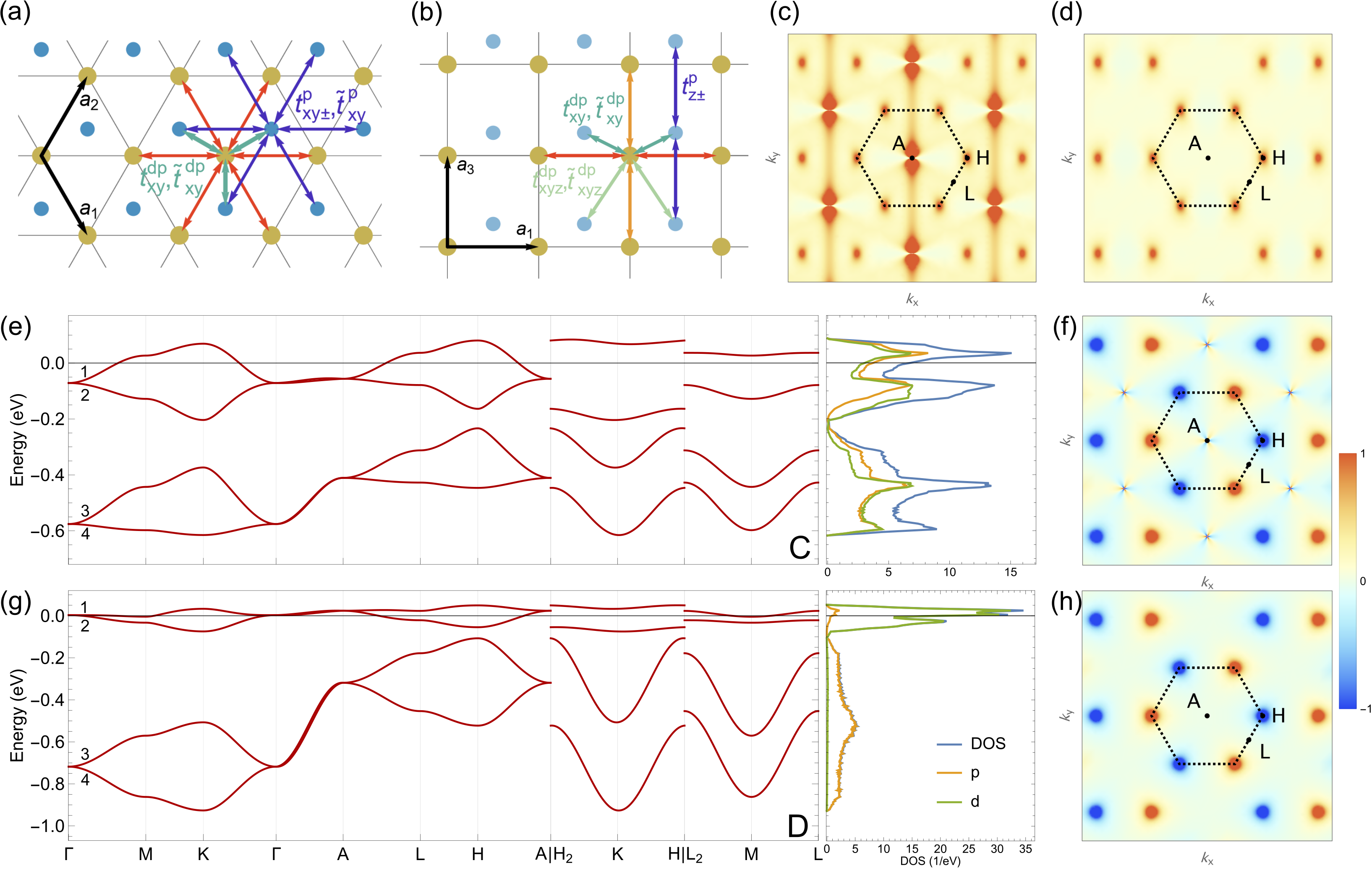}
\caption{
\textbf{Four-band model for SG~\textit{P}3}. [(a),(b)] Top and side view of the trigonal lattice with nearest- and next-nearest-neighbor hopping terms, where symmetry-related directions are shown in the same color. 
(e,g) Resulting band structures with the corresponding (orbital-resolved) density of states (DOS) for parameter sets C and D, see Table~\ref{tab:parameters}.
[(c),(d)] Quantum metric $g^{xx}$ for band 2 and bands (12)  with parameter set C.
[(f),(h)] Berry curvature $\Omega^{xy}$ for bands 2 and (12) with parameter set C. 
} \label{fig:FourBandModel}
\end{figure*}

\subsection{Two-band model}
To construct an effective model for the doping-induced states in a lattice of SG~$P6_3/m$, we assume that the relevant states form a lattice with SG~$P3$. 
Without loss of generality, they occupy the 1a Wyckoff position of the trigonal unit cell. 
We introduce all symmetry-allowed nearest-neighbor hopping terms, see Figs.~\ref{fig:TwoBandModel}(a) and \ref{fig:TwoBandModel}(b).
To capture the band topology in a symmetry-enforced fashion, we use the complex representation of the threefold rotation $C_3$, created by the operators $\{\hat d_1^\dagger,\hat d_2^\dagger\}$ corresponding to the rotation eigenvalues $\mathrm{e}^{2 \pi i/3}$ and $\mathrm{e}^{-2 \pi i/3}$, respectively. This can capture any band pair comprising $d_{xz}$-, $d_{yz}$-, $d_{x^2-y^2}$-, and $d_{xy}$-orbital weights, see Appendix~\ref{app:twobandmodel_symmetries}.
We find the Hamiltonian
\begin{align}
     H_d(\bk) &= \begin{pmatrix}H_{11}(\bk) & H_{12}(\bk) \\[1mm] H^*_{12}(\bk) & H_{22}(\bk) \end{pmatrix}
     \label{eqn:HTwoBand}
\end{align}
\vspace{-0.2cm}
with
\vspace{0.1cm}
%
{\allowdisplaybreaks %
\begin{align}
    &H_{11}(\vb{k}) = -\mu_d + t^d_{z+} \cos\big(\vb{k}\cdot \vb{a}_3\big) + t^d_{z-} \,\sin\big(\vb{k}\cdot \vb{a}_3\big)
    \nonumber
    \\
    &
    +
    t^d_{xy+} \Big[\!\cos\!\big(\vb{k}\!\cdot\! \vb{a}_1\big) \!+\! \cos\!\big(\vb{k}\!\cdot\! (\vb{a}_1 \!+ \vb{a}_2)\big) \!+ \cos\!\big(\vb{k}\!\cdot\! \vb{a}_2\big) \Big]
    \nonumber
    \\
    &
    +
    t^d_{xy-} \Big[ \!\sin\!\big(\vb{k}\!\cdot\! \vb{a}_1\big) - \sin\!\big(\vb{k}\!\cdot\! (\vb{a}_1 \!+\! \vb{a}_2)\big) + \sin\!\big(\vb{k}\!\cdot\! \vb{a}_2\big) \Big]\, , 
    \label{eqn:H11d}
    \\
    &H_{22}(\vb{k}) 
    =  H_{11}(-\vb{k}) \, ,\\
    &H_{12}(\vb{k}) 
    =
    \tilde t^d_{xy} \Big[  \cos\big(\vb{k}\cdot \vb{a}_1\big) + \mathrm{e}^{i \frac{2 \pi}{3} } \cos\big(\vb{k}\cdot (\vb{a}_1+ \vb{a}_2)\big) 
    \nonumber
    \\ 
    & \hspace{2.5cm}
    + \mathrm{e}^{-i \frac{2 \pi}{3} }  \cos\big(\vb{k}\cdot \vb{a}_2\big) \Big]
    , 
\end{align} }
%
\!\!where we use the lattice vectors $\vb{a}_1 = a (1/2, - \sqrt{3}/2, 0)$, $\vb{a}_2 = a (1/2,  \sqrt{3}/2, 0)$, and $\vb{a}_3 = c (0,0,1)$, setting $a = c = 1$. The model has six parameters, the $\vb{a}_1$-$\vb{a}_2$-plane hoppings $t^d_{xy\pm}$ and $\tilde t^d_{xy}$, the out-of-plane hoppings $t^d_{z\pm}$, and the chemical potential $\mu_d$. The subscripts denote the real space directions and whether the hopping is inversion symmetric $(+)$ or antisymmetric $(-)$. We provide longer-range hoppings and the Hamiltonian in real space in the Appendixes~\ref{app:twobandmodel_longerRange} and \ref{app:realspaceversion}.
Two band structures along the path sketched in Fig.~\ref{fig:TwoBandModel}(c) are shown in Figs.~\ref{fig:TwoBandModel}(e) and \ref{fig:TwoBandModel}(h) for different parameter sets, see Table~\ref{tab:parameters}.

The band gap vanishes at $\Gamma$ and A independent of parameters, as enforced by time-reversal symmetry. 
The gaps at M and L (K and H) are identical and scale with $|\tilde t^d_{xy}|$ ($|t^d_{xy-}|$). The splitting on $\Gamma$-A is proportional to $|t^d_{z-}|$. The comparison of the band curvatures at the band touching points $\Gamma$ and A give insight into the sign of $t^d_{xy+}$ and its relative size compared to $|\tilde t^d_{xy}|$, see Appendix~\ref{app:twobandmodel_parameterAnalysis}.

\subsection{Four-band model}
Inspired by the hybridization between the copper Cu and extra oxygen O or $(\text{OH})_2$ orbitals, as seen in DFT \cite{Griffin2023, Si2023, Lai2023, Kurleto2023}, we introduce a second site to our lattice on the 1b Wyckoff position of space group $P3$.
The resulting second band pair exhibits the complex representation of $P3$, referred to as $\{\hat p^\dagger_1, \hat p^\dagger_2\}$. This can capture any band pair comprising $p_{x}$- and $p_{y}$-orbital weights.
Our four-band model takes the form
\begin{align}
    H_{4\times4}(\bk)
    &=
    \begin{pmatrix}H_{d}(\bk) & H_{dp}(\bk) \\[1mm] H^\dagger_{dp}(\bk) & H_{p}(\bk) \end{pmatrix} \, .
    \label{eqn:HFourBand}
\end{align}
The hopping terms, see Figs.~\ref{fig:FourBandModel}(a) and \ref{fig:FourBandModel}(b), within each sublattice take the same form as in the two-band model. Thus, we keep $H_d(\bk)$ for the $d$ sublattice and define $H_p(\bk)$ for the $p$ sublattice by replacing $d \rightarrow p$ in all indices of $H_d(\bk)$ in Eq.~\eqref{eqn:HTwoBand}. We give the exact form of the inter-sublattice hopping $H_{dp}(\bk)$ in Appendix~\ref{app:fourbandmodel_hopping}, the symmetries and basis convention in Appendix~\ref{app:fourbandmodel_convention}, and the real space description in Appendix~\ref{app:fourbandmodel_realspace}. 
Two band structures are shown in Figs.~\ref{fig:FourBandModel}(e) and \ref{fig:FourBandModel}(g) for different parameter sets, see Table~\ref{tab:parameters}.

\section{Results}

\subsection{Topology and quantum geometry}

In a multiband system, not only the band dispersion but also the quantum states generically exhibit a nontrivial momentum dependence, which is described by quantum geometry locally and topology globally. We introduce the projector formalism for multiband quantum geometry and apply it to the two- and four-band models.

\subsubsection{Projector formalism for multiband quantum geometry}

We use the convenient mathematical description in terms of band projectors $P_n(\bk)=|u_n(\bk)\rangle\langle u_n(\bk)|$
\cite{Graf2021, Mera2022, Avdoshkin2023} with the defining property $P_n(\bk)P_m(\bk)=\delta_{nm}P_n(\bk)$.
The projectors are unaffected by the $U(1)$ gauge ambiguity of the cell-periodic Bloch wave function $|u_n(\bk)\rangle$ of each band $n$. 
Single-band projectors are not well defined at the momenta of band touching. 
Thus, in addition to the projectors of each band, we construct the projector on two bands isolated from the rest, i.e., $P_{(12)}(\bk)=P_1(\bk)+P_2(\bk)$. This projector on the subsystem is gauge invariant under $U(2)$ gauge transformations \cite{Mera2022}. Even if there are band crossings between the two bands, the projector $P_{(12)}$ will be well defined for all momenta.

A projector $P_j(\bk)$ on a single band or larger subspace defines the corresponding quantum metric $g^{\alpha\beta}_j(\bk)$ and Berry curvature $\Omega^{\alpha\beta}_j(\bk)$ via 
\begin{align}
    &g^{\alpha\beta}_j(\bk)=\frac{1}{2}\text{Tr}\Big[\partial_\alpha P_j(\bk)\,\partial_\beta P_j(\bk)\Big] \,, \\
    &\Omega^{\alpha\beta}_j(\bk)=i\,\text{Tr}\Big[P_j(\bk)\partial_\alpha P_j(\bk)\partial_\beta P_j(\bk)\Big]-(\alpha\leftrightarrow \beta) \, ,
\end{align}
with trace Tr and momentum derivative $\partial_\alpha=\partial_{k_\alpha}$ in direction $\alpha$. Note that the quantum metric is in general nonadditive, that is $g^{\alpha\beta}_{(12)}=g^{\alpha\beta}_1+g^{\alpha\beta}_2+\text{Tr}\big[\partial_\alpha P_1 \partial_\beta P_2\big]$ when $P_{(12)}=P_1+P_2$, where the last cross term involves the projector of both bands 1 and 2 \cite{Mera2022}. 

A two-band Hamiltonian in the Bloch basis can be written in the form $H(\bk) = d_0(\bk) + d_x(\bk) \sigma_x + d_y(\bk) \sigma_y + d_z(\bk) \sigma_z$ with Pauli matrices $\sigma_x, \sigma_y, $ and $\sigma_z$ and momentum-dependent functions $d_0(\bk), d_x(\bk), d_y(\bk), $ and $d_z(\bk)$. We do not write the identity matrix explicitly. It is useful to define $d(\bk)=\sqrt{d_x(\bk)^2+d_y(\bk)^2+d_z(\bk)^2}$. The projectors onto the two bands take the form $P_\pm(\bk)=\frac{1}{2}\big(1\pm\bn(\bk)\cdot \bsigma\big)$, where $\bn(\bk)=\big(d_x(\bk),d_y(\bk),d_z(\bk)\big)/d(\bk)$ and $\bsigma=(\sigma_x,\sigma_y,\sigma_z)$ leading to the commonly used expressions $g^{\alpha\beta}_\pm(\bk)=\frac{1}{4}\partial_\alpha \bn(\bk)\cdot \partial_\beta \bn(\bk)$ and $\Omega^{\alpha\beta}_\pm(\bk)=\mp\frac{1}{2}\bn(\bk)\cdot \big(\partial_\alpha \bn(\bk)\times \partial_\beta \bn(\bk)\big)$.
Note that the quantum metrics of both bands are identical and the Berry curvatures have opposite signs. 

The numerical evaluation of the quantum metric and Berry curvature in the projector formalism is straightforward. For a fixed momentum, we obtain the complex vector $|u_n(\bk)\rangle$ by diagonalizing $H(\bk)$ numerically and construct the respective projector matrix. The gauge invariance enables the use of discrete derivatives of the projectors, i.e., $\partial_x P_j(\bk)\approx\frac{1}{2\delta}\big[P_j(k_x+\delta,k_y,k_z)-P_j(k_x-\delta,k_y,k_z)\big]$ for sufficiently small $\delta>0$. Numerical accuracy can be checked by projector identities such as $\big(\partial_x P_n(\bk)\big)P_m(\bk)+P_n(\bk)\big(\partial_x P_m(\bk)\big)=0$ for $n\neq m$.

\subsubsection{Application to the two- and four-band models}

As expected for spinless representations \cite{tsirkin2017, alpin2023}, 
we find double Weyl points with Chern number $\nu = -2$ and $\nu = +2$ at $\Gamma$ and A, respectively, as shown in Fig.~\ref{fig:TwoBandModel}(d) \cite{fukui2005}.
We show the Berry curvature $\Omega^{xy}_2$ in Figs.~\ref{fig:TwoBandModel}(f) and \ref{fig:TwoBandModel}(i) and the quantum metric $g^{xx}_2$ in Figs.~\ref{fig:TwoBandModel}(g) and \ref{fig:TwoBandModel}(j) for $k_z=0$ and band 2, where the Weyl points lead to divergences. 
Depending on the gap size on $\Gamma$-A, see Appendix~\ref{app:twobandmodel_parameterAnalysis}, the Berry curvature in Fig.~\ref{fig:TwoBandModel}(f) is larger compared to Fig.~\ref{fig:TwoBandModel}(i). 
Whereas the quantum metric is only large around $\Gamma$ in Fig.~\ref{fig:TwoBandModel}(j), it has extended regions with larger contributions in Fig.~\ref{fig:TwoBandModel}(g). 
To illustrate that the dependence on the $\Gamma$-A gap is a generic feature, we discuss the plane $k_z = \pi/2$ in Appendix~\ref{app:twobandmodel_berryandmetric}.
More than two bands are required to see nontrivial effects of multiband geometry since the geometry of the combined bands vanishes for a two-band model, $g^{xx}_{(12)}=\Omega^{xy}_{(12)}=0$.

In Figs.~\ref{fig:FourBandModel}(c), \ref{fig:FourBandModel}(d), \ref{fig:FourBandModel}(f), and \ref{fig:FourBandModel}(h), we show the Berry curvature $\Omega^{xy}$ and quantum metric $g^{xx}$ at $k_z=\pi$ for the four-band model C, cf. Appendix~\ref{app:fourbandmodel} and Table~\ref{tab:parameters}, with band structure and density of states shown in Fig.~\ref{fig:FourBandModel}(e). Here, both $d$ and $p$ sites have the same chemical potential $\mu_d=\mu_p$, which requires a strong \mbox{d-p} coupling $H_{dp}$ in order to obtain the shown band gap. This implies a large orbital mixing between bands. Indeed, the quantum metric $g^{xx}_2$ of band~2 shows additional features at H, see Fig.~\ref{fig:FourBandModel}(c), which are not present in the two-band model. As shown in Fig.~\ref{fig:FourBandModel}(d), these features are present in the quantum metric $g^{xx}_{(12)}$ of the combined bands 1 and 2. This metric shows no singularity, as expected, because the singularities are compensated by the additional cross-term \cite{Mera2022}.
In Figs.~\ref{fig:FourBandModel}(f) and \ref{fig:FourBandModel}(h) we show the Berry curvature $\Omega^{xy}_2$ and $\Omega^{xy}_{(12)}$, respectively. Similarly to the quantum metric, we observe no divergence and find additional contributions at H due to the band gap minimum between band 2 and 3. The four-band model D, see Table~\ref{tab:parameters}, has only weak d-p coupling and does not show significant additional features compared to the two-band model.

The quantum geometry of an isolated subspace of bands is an interesting property connected to new geometric effects in narrow or flat bands, where the interaction strength exceeds the subspace band width \cite{Peotta2015, Hofmann2020, Mitscherling2022, Mao2023b}. A key quantity is the integrated quantum metric $\bar g_{(12)}\equiv \int \frac{d^3\bk}{V} \tr g_{(12)}(\bk)$ with $V=16\pi^3/\sqrt{3}$, which can be interpreted as the gauge-invariant part of the Wannier function  spread in real space \cite{Marzari2012}. We find $\bar g_{(12)}=0.26$ and $\bar g_{(12)}=0.03$ for model C and D, respectively. Geometric contributions to observables are, thus, expected to be significantly larger for model C than D. 

\subsection{Singular flat bands}

Systems with flat bands are promising candidates for phenomena that rely dominantly on quantum geometry. 
As seen in Fig.~\ref{fig:TwoBandModel}(h) the two-band model appears to exhibit a flat-band limit with non-vanishing hopping terms, which we identify in the following as a singular flat band. 

\subsubsection{Flat-band conditions for general two-band models}
\label{sec:flatbandcondition}

The upper and lower band of a generic two-band Hamiltonian take the form $E_\pm(\bk) = d_0(\bk)\pm d(\bk)$.
Using this particular form of the eigenvalues, we obtain a flat band for one of the two bands when the condition 
\begin{align}
    d(\bk)^2=d_x^2(\bk)+d_y^2(\bk)+d_z^2(\bk)=(d_0(\bk)+c_0)^2
\end{align}
is fulfilled for a momentum constant $c_0$. In this case, we have (i) $E_+(\bk)=2d_0(\bk)+c_0$ and $E_-(\bk)=-c_0$ for $d_0(\bk)+c_0>0$, and (ii) $E_+(\bk)=-c_0$ and $E_-(\bk)=2d_0(\bk)+c_0$ for $d_0(\bk)+c_0<0$.
Thus, we obtain conditions on the parameters in the Hamiltonian by enforcing 
\begin{align}
    f(\bk)=d_x^2(\bk)+d_y^2(\bk)+d_z^2(\bk)-(d_0(\bk)+c_0)^2=0
    \label{eqn:flatbandCondition}
\end{align}
for all relevant momenta $\bk$, for which the band should be flat. If we assume that $f(\bk)$ is an analytic function, which is fulfilled for tight-binding Hamiltonians, we can alternatively Taylor expand $f(\bk)$ around a specific momentum with finite band gap and solve the (generically infinite) set of equations enforcing the expansion coefficients to vanish. 
Applying this strategy to the two-band model in Eq.~\eqref{eqn:HTwoBand}, we find a flat band at $k_z=0,\pi$ for $2t^d_{xy+}=\pm |\tilde t^d_{xy}|$ and $2t^d_{xy-}=\pm \sqrt{3}|\tilde t^d_{xy}|$. The band remains flat for all $k_z$ if $t^d_{z-}=0$. 
If $t^d_{z-}\neq 0$ the flat band gets dispersive for $k_z\neq 0$ as expected due to the finite Chern number. 

\subsubsection{Compact localized state (CLS)}
\label{sec:CLS}

Without loss of generality, let us assume that the lower band is flat. Using the condition in Eq.~\eqref{eqn:flatbandCondition}, the flat-band eigenvector fulfilling $H(\bk)\uvec_\text{flat}(\bk)=-c_0\,\uvec_\text{flat}(\bk)$ takes the form
\begin{align}
    &\uvec_\text{flat}(\bk)=\frac{1}{\alpha_\bk}\begin{pmatrix} d_z(\bk)-d_0(\bk)-c_0 \\[2mm] d_x(\bk)+id_y(\bk) \end{pmatrix}
    \label{eqn:vflat}
\end{align}
with $\alpha_\bk=\sqrt{2}\sqrt{(d_0(\bk)+c_0)(d_0(\bk)+d_z(\bk)+c_0)}$, see Appendix~\ref{app:flatband}.
Following J.-W.~Rhim and B.-J.~Yang \cite{Rhim2019}, 
the corresponding CLS is given by 
\begin{align}
    |\chi_\bR\rangle = \sum_{\bR'} w^1_{\bR,\bR'}|\bR',d_1\rangle+\sum_{\bR'} w^2_{\bR,\bR'}|\bR',d_2\rangle
    \label{eqn:CLS}
\end{align}
with
\begin{align}
    \mathbf{w}_{\bR,\bR'}=\frac{c_\chi}{\sqrt{N_c}}\sum_{\bk}\alpha_\bk\, e^{i\bk\cdot(\bR'-\bR)}\uvec_\text{flat}(\bk) \, ,
    \label{eqn:CLS2}
\end{align}
where $w^\alpha_{\bR,\bR'}$ are the components of $\mathbf{w}_{\bR,\bR'}$ for the two distinct $d$-orbital states $\hat d_1$ and $\hat d_2$. $|\bR,d_\alpha\rangle$ denotes the state of $\alpha=1,2$ within the unit cell labeled by $\bR$. $N_c$ is the number of unit cells and $c_\chi$ is the normalization constant. The CLS given in Eq.~\eqref{eqn:CLS} is an eigenfunction of the flat band. Note that the translation-symmetry-related $|\chi_\bR\rangle$ for different $\bR$ may not be orthogonal. They do not necessarily form a complete basis when $\alpha_\bk=0$ for some momentum $\bk$ \cite{Rhim2019}. 

\begin{figure}[t!] 
\includegraphics[width = 0.45 \textwidth]{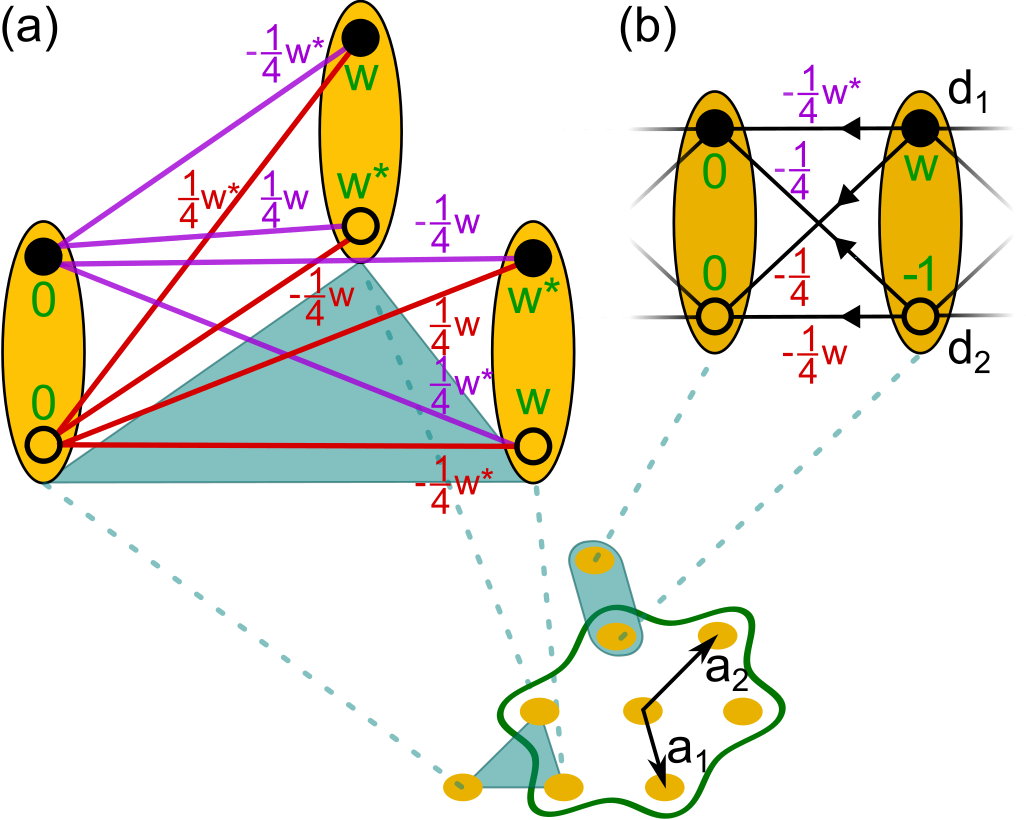}
\caption{
Visualization of the compact localized state (CLS) of the singular flat band and the destructive interference for triangles (a) and lines (b). The orbital weights (green) and hoppings (red, purple) involve $w=(1+i\sqrt{3})/2$.
} \label{fig:CreutzLadder}
\end{figure} 

We give the following concrete example by choosing  $t^d_{xy+}=-1/4$, $t^d_{xy-}=-\sqrt{3}/4$, $\tilde t^d_{xy}=-1/2$, $t^d_{z+}=-1/2$, and $\mu_d=-1/4$, where we obtain a flat lower band at energy $E_-(\bk)=-1$ for $k_z=0$ and $E_-(\bk)=0$ for $k_z=\pi$ with band gap $E_+(\bk)-E_-(\bk)=2$ at momentum $M=(\pi,-\pi/\sqrt{3},0)$. The corresponding band dispersion is shown in Appendix~\ref{app:flatband}. It resembles the band structure shown in Fig.~\ref{fig:TwoBandModel}(h).
The flat-band eigenvector at $k_z=0\,(\pi)$ constructed as described above reads
\begin{align}
    &\uvec_\text{flat}(\bk)=\frac{1}{\alpha_\bk}\Biggr[\begin{pmatrix} 
        -3 \\ 0 
    \end{pmatrix}
    +
    \begin{pmatrix} 
        w^* \\ w
    \end{pmatrix}
    e^{i\bk\cdot(\vb{a}_1+\vb{a}_2)} \nonumber
    \\ & +    
    \begin{pmatrix} 
        w \\ w
    \end{pmatrix}
    e^{-i\bk\cdot(\vb{a}_1+\vb{a}_2)}
    +
    \begin{pmatrix} 
        w \\ w^*
    \end{pmatrix}
    e^{i\bk\cdot\vb{a}_2}
    +
    \begin{pmatrix} 
        w^* \\ w^*
    \end{pmatrix}
    e^{-i\bk\cdot\vb{a}_2} \nonumber
    \\ &+
    \begin{pmatrix} 
        w \\ -1
    \end{pmatrix}
    e^{i\bk\cdot\vb{a}_1}
    +
    \begin{pmatrix} 
        w^* \\ -1
    \end{pmatrix}
    e^{-i\bk\cdot\vb{a}_1}\Biggr]
    \label{eqn:CLSModel}
\end{align}
with $w=\frac{1}{2}(1+i\sqrt{3})=-e^{-i2\pi/3}$ and its complex conjugate $w^*$.  Using Eqs.~\eqref{eqn:CLS} and \eqref{eqn:CLS2}, we can read off the CLS. One of these nonorthogonal CLS involves a site and its six nearest neighbors in the $x$-$y$ plane, see Fig.~\ref{fig:CreutzLadder}. The localization crucially relies on the destructive interference between the two orbitals as sketched in Fig.~\ref{fig:CreutzLadder}(a) and \ref{fig:CreutzLadder}(b).
This destructive interference via intra- and interorbital hopping resembles the $1d$ Creutz ladder \cite{Creutz1999}.
Since the normalization $\alpha_\bk$ vanishes at $\Gamma=(0,0,0)$, see Appendix~\ref{app:flatband}, and lifting the band crossing leads to a Chern band for $t^d_{z-}\neq 0$, the flat bands for $k_z= 0$ and $k_z=\pi$ are {\it singular} and the set of all translation-related CLS do not form a complete basis \cite{Rhim2019}.


\section{Discussion}

\subsection{Application to pristine copper-doped lead apatite}
Our band structures shown in Fig.~\ref{fig:TwoBandModel}(e), \ref{fig:TwoBandModel}(h), \ref{fig:FourBandModel}(e), and \ref{fig:FourBandModel}(g) are able to reproduce most of the key features reported in DFT studies on copper-doped lead apatite that assumed the same regular dopant placement \cite{Griffin2023, Si2023, Lai2023, Kurleto2023}. For the two-band model we have identified two parameter sets A and B, see Table~\ref{tab:parameters}, to capture the band structure for different relative positions of copper and extra oxygen sites \cite{Lai2023}. 
Parameter set B generates the reported almost flat band \cite{Griffin2023, Lai2023, Tavakol2023}. Our models show saddle-point Van Hove singularities in all bands at M and L with an energy profile as reported in Ref.~\cite{Lai2023}. 

The four-band model with parameter set C, see Table~\ref{tab:parameters}, is constructed to reproduce the nearly equal distribution of $p$ and $d$ orbitals in the density of states, shown in Fig.~\ref{fig:FourBandModel}(e), which was seen in the DFT results \cite{Griffin2023, Si2023, Lai2023, Kurleto2023} for bands 1 and 2. The weak d character of bands 3 and 4 in the DFT results cannot be captured by just four bands. Model D corresponds to a weakly coupled four-band model with mixing of d and p orbitals of at most 10\%.

Later DFT studies include tight-binding models with parameters reasonably close to ours \cite{Jiang2023a, Fidrysiak2023, Si2023b, Mao2023c, Yue2023, Bai2023, Korotin2023, Witt2023}. Note that these models use cubic harmonics, whereas our models correspond to $C_3$-symmetric combinations of $d$ and $p$ orbitals, see Appendix~\ref{app:twobandmodel_symmetries}.
The topology and quantum geometry analyzed in Ref.~\cite{Jiang2023a} align with our conclusions.

We emphasize that there is a discrepancy between the assumed pristine material structure in DFT and the materials studied experimentally.  
Copper-doped lead apatite $\text{Pb}_{9}\text{Cu}(\text{PO}_4)_6\text{O}$ samples were found to exhibit comparable substitution on both $4f$ and $6h$ Wyckoff positions as well as a clustering of copper~\cite{Puphal2023}. 
Since the site symmetry of $6h$ only comprises mirror symmetry, no complex representation exists, and therefore a metallic normal state in the absence of interactions, as described in the present paper, is not expected for these samples.
The originally proposed unusual resistivity signatures of copper-doped lead apatite (referred to as \mbox{"LK-99"}) \cite{Lee2023, Lee2023a} have been connected to a metal insulator transition \cite{Lee2023b}, which was experimentally related to $\text{CuS}_2$ impurities \cite{Jiang2023a, Jain2023, Puphal2023, Zhu2023}.

\subsection{Further candidate materials}
Our results apply to a range of doped variants of established semiconductors. 
In the rhombohedral van der Waals ferromagnet Cr$_2$Ge$_2$Te$_6$ \cite{Gong2017, Verzhbitskiy2020} Mn doping increases the carrier density, which coincides with the appearance of a complex representation of the $C_3$ symmetry, as described in Eq.~\eqref{eqn:HTwoBand}, and is found to improve the thermoelectric figure of merit \cite{Tang2017}.
Doped monolayers of the well-studied semiconductor SiC \cite{Kimoto2014, Roccaforte2021} have in-gap states with bands resembling models A or B at $k_z =0$, dependent on the chosen dopant \cite{Luo2018}. 
In the 2D dichalcogenide MoS$_2$ sulfur vacancies act as active centers for hydrogen evolution catalysis \cite{Cao2021} and, interestingly, also lead to an impurity band that is qualitatively described by model A \cite{Han2019, Loh2021}.

\begin{table}[t!]
    \centering
    \setlength{\arrayrulewidth}{0.2mm}
    \begin{tabular}{l c c c c}
        \hline \hline
          & 2-band A \hspace{0.4cm} & 2-band B \hspace{0.4cm} & 4-band C \hspace{0.4cm} & 4-band D \\[1mm]\hline
          $\mu_d$               & $10 $     & $0$       & $250$     & $20$      \\[1mm]\hline
          $t^d_{z+}$            & $-30$     & $-8$      & $-20$     & $-10$    \\[1mm]\hline
          $t^d_{z-}$            & $5$       & $2$       & $-3$      & $-2.5$    \\[1mm]\hline
          $t^d_{xy+}$           & $-5$      & $-4$      & $-20$     & $-5$  \\[1mm]\hline
          $t^d_{xy-}$           & $30$      & $5$       & $-14$     & $20$   \\[1mm]\hline
          $\tilde t^d_{xy}$     & $15$      & $6$       & $14$      & $10$    \\[1mm]\hline
          $\mu_p$               &           &           & $250$     & $500$      \\[1mm]\hline
          $t^p_{z+}$            &           &           & $-70$     & $-200$     \\[1mm]\hline
          $t^p_{z-}$            &           &           & $8$       & $10$    \\[1mm]\hline
          $t^p_{xy+}$           &           &           & $1$       & $10$    \\[1mm]\hline
          $t^p_{xy-}$           &           &           & $-20$     & $80$      \\[1mm]\hline
          $\tilde t^p_{xy}$     &           &           & $-14$     & $70$     \\[1mm]\hline
          $t^{dp}_{xy}$         &           &           & $70$      & $50$      \\[1mm]\hline
          $\tilde t^{dp}_{xy}$  &           &           & $100$     & $10$     \\[1mm]\hline
          $t^{dp}_{xyz}$        &           &           & $14$      & $10$      \\[1mm]\hline
          $\tilde t^{dp}_{xyz}$ &           &           & $14$      & $10$
        \\[1mm]\hline \hline
    \end{tabular}
    \caption{Summary of the used parameters to capture approximately the key features of the DFT results in Refs.~\cite{Griffin2023, Si2023, Lai2023, Kurleto2023}. All units are meV.
    }
    \label{tab:parameters}
\end{table}

\section{Conclusion}
We discussed how doping-induced states in semiconductors and insulators enable the design of topological and geometrical band structures by choosing a suitable site symmetry.
The simplicity yet richness of the constructed two- and four-band model makes them an optimal starting point to study the implications of the discussed nontrivial multiband quantum geometry and singular flat bands. 
Analogously to the presented example, our method can be applied to other dopant configurations and other doped insulating materials \cite{Tang2017, Luo2018, Loh2021, Georgescu2023}, where symmetry-enforced band crossings occur whenever a complex representation exists due a three-, four-, or sixfold rotation symmetry.


\begin{acknowledgements}
The authors thank Alexander Avdoshkin, Zhehao Dai, Tomohiro Soejima, Jan~M.~Tomczak, and Taige Wang for very stimulating discussions.
M.M.H.~is funded by the Deutsche Forschungsgemeinschaft (DFG, German Research Foundation) - project number 518238332.
J.M. acknowledges support by the German National Academy of Sciences Leopoldina through Grant No. LPDS 2022-06.
\end{acknowledgements}

\appendix

\section{Two-band model}
\label{app:twobandmodel}

\subsection{Symmetry operations}
\label{app:twobandmodel_symmetries}

The orbital character of our basis states is introduced by the imposed symmetries. 
We represent space group $P3$ comprising $C_3$ rotation and time reversal $\theta$ by 
\begin{align}
    U^d_{C_3} = \begin{pmatrix} \mathrm{e}^{i \frac{2 \pi}{3} } & 0 \\[1mm] 0 & \mathrm{e}^{-i \frac{2 \pi}{3} } \end{pmatrix} 
    \text{ and }
    \theta = \begin{pmatrix} 0 & 1 \\[1mm] 1 & 0 \end{pmatrix} K,
    \label{eqn:twoBandSymmetries}
\end{align}
respectively, where $K$ refers to the complex conjugation operator. 
With the above definitions our two-band model in Eq.~\eqref{eqn:HTwoBand} fulfills 
\begin{align}
    &(U^{d}_{C_3})^\dagger H(\bk) U^{d}_{C_3} = H_d(C_3^{-1} \bk)\, ,\\
    &\theta H_d(\bk) \theta = H_d(- \bk)\,, 
\end{align}
where  $C_3$ is the conventional action on spatial coordinates,
\begin{align}
    C_3 = \mqty( \cos(2\pi/3) & -\sin(2\pi/3) \\ \sin(2\pi/3) & \cos(2\pi/3) ).
\end{align}
This complex representation can model any $C_3$-symmetric superposition of orbitals such as $d$ orbitals. 
If these orbitals are expressed in spherical harmonics $Y_l^m(\theta, \varphi)$, the quantum number $m$ and the eigenvalues $\lambda_{C_3}$ of threefold rotation fulfill $\lambda_{C_3} = \exp(i m 2\pi /3)$. 
Thus, the first (second) component of our basis can represent any superposition of states described by $m=1,-2$ ($m=-1,2$).
Expressed in cubic harmonics, the complex representation may exhibit contributions from ($d_{xz}$, $d_{yz}$) and ($d_{xy}$, $d_{x^2-y^2}$), which comprise states with $|m|=1$ and $|m|=2$, respectively.

\subsection{Possible additional symmetry-allowed hopping terms}
\label{app:twobandmodel_longerRange}

In the main text, we have considered only nearest-neighbor hopping terms. 
To fit the model to a specific band structure, we also provide second- and third-nearest-neighbor hopping terms, with distances in real space $\sqrt{2}$ and $\sqrt{3}$ in units of the lattice constant, respectively. 
The second-nearest-neighbors that occur are the six vectors shown in Fig.~\ref{fig:TwoBandModel}(a) combined with a step in the $\vb{a}_3$ direction, which result in a Hamiltonian $\delta H_{11}(\bk)$ that can be added to $H_{11}(\bk)$ in Eq.~\eqref{eqn:H11d}. 
Due to the absence of mirror or additional rotation symmetries in the site symmetry, this corresponds to six generally different terms in $\delta H_{11}(\bk)$, see Eqs.~\eqref{eqn:deltaH11} and \eqref{eqn:deltaH22} below.
The terms in $\delta H_{11}(\bk)$ can be used to describe anisotropy along the $k_z$ direction, e.g., the line H$_2$-K-H similar to $t_{z-}^d$.
But unlike $t_{z-}^d$ the longer range $\delta H_{11}(\bk)$ acts differently on the axes $\Gamma$-A and H$_2$-K-H. 
The third-nearest neighbor terms $\delta H_{12}(\bk)$ are given in  Eq.~\eqref{eqn:deltaH12} below,
\begin{widetext}
{\allowdisplaybreaks
\begin{align}
    \delta H_{11}(\bk)
    &= 
    t_{1,xyz+}^d \Big(\cos\big( \bk \cdot  (\vb{a}_1 + \vb{a}_3)\big) + \cos\big( \bk \cdot  (-\vb{a}_1 - \vb{a}_2 + \vb{a}_3)\big) + \cos\big( \bk \cdot  (\vb{a}_2 + \vb{a}_3) \big) \,\Big) 
    \nonumber
    \\
    &+
    t_{2,xyz+}^d \Big(\cos\big( \bk \cdot  (\vb{a}_1 - \vb{a}_3)\big) + \cos\big( \bk \cdot  (-\vb{a}_1 - \vb{a}_2 - \vb{a}_3)\big) + \cos\big( \bk \cdot  (\vb{a}_2 - \vb{a}_3)\big)  \,\Big) 
    \nonumber
    \\
    &+ 
    t_{1,xyz-}^d \Big(\sin\big( \bk \cdot (\vb{a}_1+ \vb{a}_3)\big) + \sin\big( \bk \cdot (-\vb{a}_1 - \vb{a}_2+ \vb{a}_3)\big) + \sin\big( \bk \cdot (\vb{a}_2+ \vb{a}_3)\big) \,\Big) 
    \nonumber
    \\
    &+ 
    t_{2,xyz-}^d \Big(\sin\big( \bk \cdot (\vb{a}_1+\vb{a}_2+ \vb{a}_3)\big) + \sin\big( \bk \cdot (-\vb{a}_2+ \vb{a}_3)\big) + \sin\big( \bk \cdot (-\vb{a}_1+ \vb{a}_3)\big)  \,\Big)
    \nonumber
    \\
    &+ t_{3,xyz-}^d \Big(\sin\big( \bk \cdot (\vb{a}_1- \vb{a}_3)\big) + \sin\big( \bk \cdot (-\vb{a}_1 - \vb{a}_2- \vb{a}_3)\big) + \sin\big( \bk \cdot (\vb{a}_2- \vb{a}_3)\big) \,\Big) 
    \nonumber
    \\
    &+ t_{4,xyz-}^d \Big(\sin\big( \bk \cdot (\vb{a}_1+\vb{a}_2- \vb{a}_3)\big) + \sin\big( \bk \cdot (-\vb{a}_2- \vb{a}_3)\big) + \sin\big( \bk \cdot (-\vb{a}_1- \vb{a}_3)\big)  \,\Big) \, ,
    \label{eqn:deltaH11}
    \\
    \delta H_{22}(\bk) 
    &= \delta H_{11}(-\bk) \, ,
    \label{eqn:deltaH22}
    \\[0.3cm]
    \delta H_{12}(\bk)
    &=
    \tilde{t}_{1,xy}^d \Big( \cos\big( \bk \cdot (2 \vb{a}_1 + \vb{a}_2)\big) + \mathrm{e}^{i \frac{2 \pi}{3} } \cos\big( \bk \cdot (-\vb{a}_1 - 2 \vb{a}_2)\big)  + \mathrm{e}^{-i \frac{2 \pi}{3} } \cos\big( \bk \cdot (- \vb{a}_1 + \vb{a}_2)\big)  \,\Big) 
    \nonumber
    \\
    &+
    \tilde{t}_{2,xy}^d \Big( \cos\big( \bk \cdot (\vb{a}_1 + 2 \vb{a}_2)\big) + \mathrm{e}^{i \frac{2 \pi}{3} } \cos\big( \bk \cdot (\vb{a}_1 - \vb{a}_2)\big)  +  \mathrm{e}^{-i \frac{2 \pi}{3} } \cos\big( \bk \cdot (-2 \vb{a}_1 - \vb{a}_2)\big)  \,\Big) \, . 
    \label{eqn:deltaH12}
\end{align}
}
%

\subsection{Real-space version of the tight-binding Hamiltonian}
\label{app:realspaceversion}

We Fourier transform the tight-binding Hamiltonian for the two-band model in Eq.~\eqref{eqn:HTwoBand} to real space via
\begin{align}
    \hat d_{\bk,\nu}=\frac{1}{\sqrt{N_c}}\sum_j \hat d_{j,\nu}e^{-i\bk\cdot(\bR_j+\brho_d)}
    \label{eqn:fourier}
\end{align}
with lattice vector $\bR_{j_1,j_2,j_3}=j_1 \vb{a}_1+j_2\vb{a}_2+j_3\vb{a}_3$, total number of unit cells $N_c$, and annihilation operator $\hat d_\nu$ of the orbital $\nu=1,2$ at the unit-cell position $\brho_d=(0,0,0)$. 
%
We obtain
\begin{align}
    \hat H = \hat H^{\phantom{\dagger}}_{11}+\hat H^{\phantom{\dagger}}_{22} + \hat H^{\phantom{\dagger}}_{12} + \hat H^\dagger_{12} - \mu_d \sum_j \big(\hat d^\dagger_{j,1}\hat d^{\phantom{\dagger}}_{j,1} +\hat d^\dagger_{j,2}\hat d^{\phantom{\dagger}}_{j,2} \big)
    \label{eqn:fourierTwoBand}
\end{align}
with
\begin{align}
    &\hat H^{\phantom{\dagger}}_{11}=\sum_j \Bigg[\,\,\,\,\frac{t^d_{xy}}{2}\,\,\,\,\Big(
    \hat d^\dagger_{j+j_1,1}\hat d^{\phantom{\dagger}}_{j,1}
    +\hat d^\dagger_{j,1} \hat d^{\phantom{\dagger}}_{j+j_1+j_2,1}
    +\hat d^\dagger_{j+j_2,1}\hat d^{\phantom{\dagger}}_{j,1}\Big)
    +\,\,\,\,t^{\phantom{d}}_z \,\,\,\,\,\,\hat d^\dagger_{j+j_3,1}\hat d^{\phantom{\dagger}}_{j,1}\Bigg]
    + \text{H.c.}\, ,\\
    &\hat H^{\phantom{\dagger}}_{22}=\sum_j \Bigg[\frac{\big(t^d_{xy}\big)^*}{2}\Big(
    \hat d^\dagger_{j+j_1,2}\hat d^{\phantom{\dagger}}_{j,2}
    +\hat d^\dagger_{j,2} \hat d^{\phantom{\dagger}}_{j+j_1+j_2,2}
    +\hat d^\dagger_{j+j_2,2}\hat d^{\phantom{\dagger}}_{j,2}\Big)
    +\big(t^{\phantom{d}}_z\big)^* \,\,\hat d^\dagger_{j+j_3,2}\hat d^{\phantom{\dagger}}_{j,2}\Bigg]
    + \text{H.c.}\, , \, 
\end{align}
with $t^d_{xy}=t^d_{xy,+}+i t^d_{xy,-}$ and Hermitian conjugate $\text{H.c.}$ as well as
\begin{align}
    \hat H_{12}=\sum_j\,\frac{\tilde t^d_{xy}}{2}\Big[
    &\big(\hat d^\dagger_{j+j_1,1}\hat d^{\phantom{\dagger}}_{j,2}+\hat d^\dagger_{j,1}\hat d^{\phantom{\dagger}}_{j+j_1,2}\big)
    +e^{i\frac{2\pi}{3}}\big(\hat d^\dagger_{j+j_1+j_2,1}\hat d^{\phantom{\dagger}}_{j,2}+\hat d^\dagger_{j,1}\hat d^{\phantom{\dagger}}_{j+j_1+j_2,2}\big)
    +e^{-i\frac{2\pi}{3}}\big(\hat d^\dagger_{j,1}\hat d^{\phantom{\dagger}}_{j+j_2,2}
    +\hat d^\dagger_{j+j_2,1}\hat d^{\phantom{\dagger}}_{j,2}\big)
    \Big]\, ,
\end{align}
and $\hat H_{21}=(\hat H_{12})^*$.
\end{widetext}

\subsection{Effect of the parameters on several model properties}
\label{app:twobandmodel_parameterAnalysis}

We give several relations between basic band structure properties and the five model parameters, the $\vb{a}_1$-$\vb{a}_2$-plane hoppings $t^d_{xy\pm}$ and $\tilde t^d_{xy}$ and the out-of-plane hoppings $t^d_{z\pm}$. Note that $\tilde t^d_{xy}$ is in general complex, whereas the other parameters are real. We focus on the high-symmetry points $\Gamma=(0,0,0)$, $M=(\pi,-\pi/\sqrt{3},0)$, $K=(4\pi/3,0,0)$, $A=(0,0,\pi)$, $L=(\pi,-\pi/\sqrt{3},\pi)$, and $H=(4\pi/3,0,\pi)$.

\subsubsection{Size of direct band gaps at high-symmetry points and on the $\Gamma$-A line}

We define the size of the direct band gaps as $\Delta_\bk=E_+(\bk)-E_-(\bk)$ and obtain
\begin{align}
    &\Delta_\Gamma=0 \,, \hspace{0.2cm}
    &&\Delta_M=4|\tilde t^d_{xy}|\, , \hspace{0.2cm}
    &&\Delta_K=3\sqrt{3}|t^d_{xy-}|\, , \hspace{0.2cm} 
    \\
    &\Delta_A=0\, , \hspace{0.2cm}
    &&\Delta_L=4|\tilde t^d_{xy}|\, , \hspace{0.2cm}
    &&\Delta_H=3\sqrt{3}|t^d_{xy-}|\, .
\end{align}
We see that $\Delta_\Gamma=\Delta_A=0$ as expected. Furthermore, $\Delta_M=\Delta_L$ and $\Delta_K=\Delta_H$. The splitting on the symmetry line $\Gamma$-A, that is $\bk=(0,0,k_z)$ is given by 
\begin{align}
    \Delta_{\Gamma\text{-}A}=2|t^d_{z-}\sin(\bk\cdot \vb{a}_3)| \, ,
\end{align}
with maximal value $\Delta^\text{max}_{\Gamma\text{-}A}=2|t^d_{z-}|$ for $k_z=\pi/2$.

\subsubsection{Band curvature at $\Gamma$ and A in various directions}

The momentum expansion of the two bands at $\Gamma=(0,0,0)$ in direction $\vb{v_{\Gamma\text{-}M}}=(M-\Gamma)/|M-\Gamma|=(\sqrt{3}/2,-1/2,0)$ and $\vb{v_{G\text{-}K}}=(K-\Gamma)/|K-\Gamma|=(1,0,0)$ reads
{
\allowdisplaybreaks
\begin{align}
    &E_+\big(k\,\vb{v_{\Gamma\text{-}M}}\big)=E_+\big(k\,\vb{v_{\Gamma\text{-}K}}\big)\nonumber\\&=\Big(3 t^d_{xy+}+t^d_{z+}-\mu_d\Big)+\frac{3}{8}\Big(|\tilde t^d_{xy}|-2t^d_{xy+}\Big)\,k^2+\mathcal{O}(k^3) \, ,\\
    &E_-\big(k\,\vb{v_{\Gamma\text{-}M}}\big)=E_-\big(k\,\vb{v_{\Gamma\text{-}K}}\big)\nonumber\\&=\Big(3 t^d_{xy+}+t^d_{z+}-\mu_d\Big)-\frac{3}{8}\Big(|\tilde t^d_{xy}|+2t^d_{xy+}\Big)\,k^2+\mathcal{O}(k^3) \, .
\end{align}
}
The momentum expansion of the two bands at $A=(0,0,\pi)$ in direction $\vb{v_{A\text{-}L}}=(L-A)/|L-A|=(\sqrt{3}/2,-1/2,0)$ and $\vb{v_{A\text{-}H}}=(H-A)/|H-A|=(1,0,0)$ reads
\begin{align}
    &E_+\big(k\,\vb{v_{A\text{-}L}}\big)=E_+\big(k\,\vb{v_{A\text{-}H}}\big)\nonumber\\&=\Big(3 t^d_{xy+}-t^d_{z+}-\mu_d\Big)+\frac{3}{8}\Big(|\tilde t^d_{xy}|-2t^d_{xy+}\Big)\,k^2+\mathcal{O}(k^3)\, ,\\
    &E_-\big(k\,\vb{v_{A\text{-}L}}\big)=E_-\big(k\,\vb{v_{A\text{-}H}}\big)\nonumber\\&=\Big(3 t^d_{xy+}-t^d_{z+}-\mu_d\Big)-\frac{3}{8}\Big(|\tilde t^d_{xy}|+2t^d_{xy+}\Big)\,k^2+\mathcal{O}(k^3)\, .
\end{align}
We read off six different scenarios. If $|\tilde t^d_{xy}|=2|t^d_{xy+}|$ or $|\tilde t^d_{xy}|=-2|t^d_{xy+}|$, the upper or the lower band remains flat. We conclude the following trends: (i) $E_+$ has positive curvature if $|\tilde t^d_{xy}|>2t^d_{xy+}$, (ii) $E_+$ has negative curvature if $|\tilde t^d_{xy}|<2t^d_{xy+}$, (iii) $E_-$ has positive curvature if $|\tilde t^d_{xy}|<-2t^d_{xy+}$, and (iv) $E_-$ has negative curvature if $|\tilde t^d_{xy}|>-2t^d_{xy+}$. 
We see that the trends depend on the relative size of $|\tilde t^d_{xy}|$ and $|t^d_{xy+}|$ as well as the sign of $t^d_{xy+}$. We summarize the result in Table~\ref{tab:CurvatureBands}.

\subsubsection{Lower-band shift at K and H with respect to $\Gamma$}

The energy of the lower band at high-symmetry points K and H with respect to $\Gamma$ is given by
\begin{align}
    &E_-(K)-E_-(\Gamma)=-\frac{3\sqrt{3}}{2}|t^d_{xy-}|-\frac{9}{2}t^d_{xy+} \, ,\\
    &E_-(H)-E_-(\Gamma)=-\frac{3\sqrt{3}}{2}|t^d_{xy-}|-\frac{9}{2}t^d_{xy+}-2t^d_{z+} \, .
\end{align}

\subsection{Berry curvature and quantum metric for the two-band model at $k_z =\pi/2$}
\label{app:twobandmodel_berryandmetric}

\begin{figure}[t!] 
\includegraphics[width = 0.45 \textwidth]{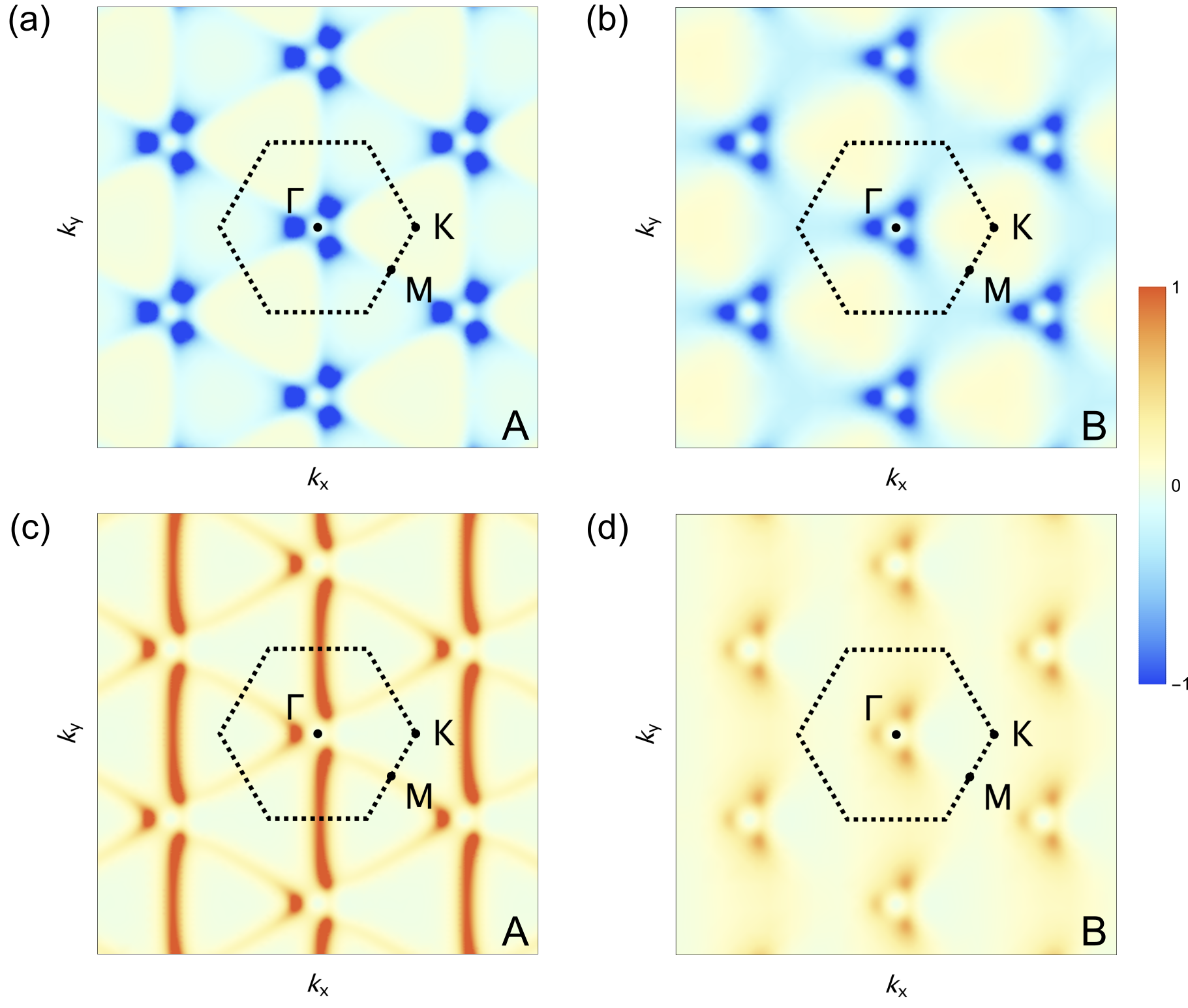}
\caption{
Berry curvature $\Omega^{xy}_-$ [(a),(b)] and quantum metric $g^{xx}_-$ [(c),(d)] at $k_z = \pi/2$ for the parameter sets A and B of the two-band model.
} \label{fig:TwoBandSupplBerryMetric}
\end{figure}

For the parameter sets A and B, see Table~\ref{tab:parameters}, we show the Berry curvature and quantum metric at $k_z = \pi/2$ in Fig.~\ref{fig:TwoBandSupplBerryMetric}.
Besides the divergence of Berry curvature and quantum metric at the Weyl points, both quantities exhibit larger values for the parameter set A, i.e., the one with the smaller gap on $\Gamma$-A.
Notably, albeit the localization of the weight in k space strongly differs between models A and B, as seen when comparing Fig.~\ref{fig:TwoBandSupplBerryMetric}(c) to Fig.~\ref{fig:TwoBandSupplBerryMetric}(d), the integrated values are of the same order of magnitude. 
Specifically, integrating the metric on the plane shown in Fig.~\ref{fig:TwoBandSupplBerryMetric}~(c)[(d)] yields $\int \frac{d^2\bk}{A} g^{xx}_-(\bk)=0.80 \,\,[0.74]$ with $A=8\pi^2/\sqrt{3}$.
For the Berry curvature the values are identical, because the integral over the plane shown in Fig.~\ref{fig:TwoBandSupplBerryMetric}(a) and \ref{fig:TwoBandSupplBerryMetric}(b) equals the Chern number of $\nu = -1$.

\section{Four-band model}
\label{app:fourbandmodel}


\subsection{Sublattice hopping in the four-band model}
\label{app:fourbandmodel_hopping}

We give the explicit form of the sublattice hopping of the four-band model in Eq.~\eqref{eqn:HFourBand}. They are
\begin{align}
    H_{dp}(\bk)
    =
    \mathrm{e}^{i \bk \cdot \big((\vb{a}_1 + 2\vb{a}_2)/3 + z \,\vb{a}_3\big)}
    \begin{pmatrix}H^{dp}_{11}(\bk) & H^{dp}_{12}(\bk) \\[1mm] H^{dp}_{21}(\bk) & H^{dp}_{22}(\bk) \end{pmatrix},
    \label{eqn:HdpFourband}
\end{align}
with the 1b Wyckoff position $ (\vb{a}_1 + 2\vb{a}_2)/3 + z \,\vb{a}_3$ and 
\begin{align} 
    H^{dp}_{11}(\bk) 
    &=
    (t^{dp}_{xy} \!+\! t^{dp}_{xyz} \mathrm{e}^{i \bk \cdot \vb{a}_3} ) 
    \Big[ 1 \!+\! \mathrm{e}^{-i \bk \cdot \vb{a}_2} \!+\! \mathrm{e}^{-i \bk \cdot (\vb{a}_1 + \vb{a}_2)} \Big] \,,
    \\
    H^{dp}_{22}(\bk) 
    &=
    H^{dp}_{11}(-\bk)^* \,, 
    \\
    H^{dp}_{12}(\bk)
    &=
    \left(\tilde{t}^{dp}_{xy} + \tilde{t}^{dp}_{xyz} \mathrm{e}^{i \bk \cdot \vb{a}_3} \right) 
    \nonumber
    \\
    &\,\,\,\times \Big[ 
    1 
    + \mathrm{e}^{i \frac{2 \pi}{3} } \mathrm{e}^{-i \bk \cdot \vb{a}_2} 
    + \mathrm{e}^{-i \frac{2 \pi}{3} } \mathrm{e}^{-i \bk \cdot (\vb{a}_1 + \vb{a}_2)} \Big] \,,
    \\
    H^{dp}_{21}(\bk) 
    &=
    H^{dp}_{12}(-\bk)^* \,.
\end{align}
Note that the relative height difference between the 1a and 1b Wyckoff positions, denoted as $z$ in Eq.~\eqref{eqn:HdpFourband}, does not affect the band structure. 
For the present paper the exact value of $z$ does not affect the results, because all considered geometric quantities are restricted to derivatives in the $k_x$-$k_y$ plane. 

\begin{table}[t!]
    \centering
    \setlength{\arrayrulewidth}{0.2mm}
    \begin{tabular}{c c c}
        \hline \hline
         $E_+$ & $E_-$ & Conditions \\[1mm]\hline
        $\nearrow$ & $\nearrow$ & $2|t^d_{xy+}|>|\tilde t^d_{xy}|$\,\,\,and\,\,\,$t^d_{xy+}<0$\\[2mm]
        $\searrow$ & $\searrow$ & $2|t^d_{xy+}|>|\tilde t^d_{xy}|$\,\,\,and\,\,\,$t^d_{xy+}>0$ \\[2mm]
        $\nearrow$ & $\searrow$ &  $|\tilde t^d_{xy}|>2|t^d_{xy+}|$
        \\[1mm]\hline \hline
    \end{tabular}
    \caption{The parameter dependence of the curvature trends at $\Gamma$ and A in direction $M, K$ and $L, H$, respectively. $\nearrow$ indicates a positive curvature. $\searrow$ indicates a negative curvature. 
    The missing combination ($\searrow$ $\nearrow$) is inconsistent with the fixed definition of the lower and upper band and, thus, left out. 
    }
    \label{tab:CurvatureBands}
\end{table}

\subsection{Symmetries and basis convention}
\label{app:fourbandmodel_convention}

The second pair of bands described by $H_p(\bk)$ obeys by itself the same symmetry representation as $H_d(\bk)$, but since the 1b Wyckoff position is not invariant under $C_3$, the application of $C_3$ moves the site into a neighboring unit cell. 
Thus, the unitary representation $U_{C_3}$ contains a factor of $\mathrm{e}^{i \bk \cdot (\vb{a}_1 + \vb{a}_2)}$. Yet, once represented in the basis convention introduced by the Fourier transform in Eq.~\eqref{eqn:fourier}, this phase cancels and one finds
\begin{align}
    U^p_{C_3} =  \begin{pmatrix} \mathrm{e}^{i \frac{2 \pi}{3} } & 0 \\[1mm] 0 & \mathrm{e}^{-i \frac{2 \pi}{3} } \end{pmatrix} 
\end{align}
leading to the definition 
\begin{align}
    U_{C_3} = \begin{pmatrix} U^d_{C_3} & 0 \\[1mm] 0 & U^p_{C_3} \end{pmatrix},
\end{align}
which is then a symmetry of the four-band model $H_{4\times4}(\bk)$ given in Eq.~\eqref{eqn:HFourBand}.
Time-reversal symmetry is local in real space, and hence is the same for the states corresponding to 1a and 1b Wyckoff positions.

\subsection{Real-space version of the tight-binding Hamiltonian}
\label{app:fourbandmodel_realspace}

The tight-binding Hamiltonian defined in Eq.~\eqref{eqn:HFourBand} is Fourier transformed to real space via Eq.~\eqref{eqn:fourier}. The annihilation operators for the $d$ and $p$ orbitals are denoted by $\hat d_{\nu}$ and $\hat p_{\nu}$ with $\nu=1,2$. The d and p orbitals are located at $\brho_d=(0,0,0)$ and $\brho_p=(\vb{a}_1+2\vb{a}_2)/3+z\vb{a}_3=(1/2, 1/2\sqrt{3}, z)$ in the unit cell. The Fourier transform of $H_d(\bk)$ is already given in Eq.~\eqref{eqn:fourierTwoBand}. The Fourier transform of $H_p(\bk)$ is equivalent to those of $H_d(\bk)$ with $d$ replaced by $p$. We give the remaining expressions of $H_{dp}(\bk)$, 
{\allowdisplaybreaks
\begin{align}
    &\hat H^{dp}_{11}\!=\!\sum_j\!\Bigg[t^{dp}_{xy}\,\Big(\hat d^\dagger_{j,1}\hat p^{\phantom{\dagger}}_{j,1}\!+\!\hat d^\dagger_{j+j_1+j_2,1} \hat p^{\phantom{\dagger}}_{j,1}\!+\!\hat d^\dagger_{j+j_2,1} \hat p^{\phantom{\dagger}}_{j,1}\Big)\nonumber\\&+\!t^{dp}_{xyz}\,\Big(\hat d^\dagger_{j,1} \hat p^{\phantom{\dagger}}_{j+j_3,1}\!+\!\hat d^\dagger_{j+j_1+j_2,1} \hat p^{\phantom{\dagger}}_{j+j_3,1}\nonumber\\&\hspace{4cm}+\hat d^\dagger_{j+j_2,1} \hat p^{\phantom{\dagger}}_{j+j_3,1})\Big)\!\Bigg]\,,\\
    &\hat H^{dp}_{22}\!=\!\sum_j\!\Bigg[\!\big(t^{dp}_{xy}\big)^*\Big(\hat d^\dagger_{j,2}\hat p^{\phantom{\dagger}}_{j,2}\!+\!\hat d^\dagger_{j+j_1+j_2,2} \hat p^{\phantom{\dagger}}_{j,2}\!+\!\hat d^\dagger_{j+j_2,2} \hat p^{\phantom{\dagger}}_{j,2}\Big)\nonumber\\&+\!\big(t^{dp}_{xyz}\big)^*\,\Big(\hat d^\dagger_{j,2} \hat p^{\phantom{\dagger}}_{j+j_3,2}+\hat d^\dagger_{j+j_1+j_2,2} \hat p^{\phantom{\dagger}}_{j+j_3,2}\nonumber\\&\hspace{4cm}+\hat d^\dagger_{j+j_2,2} \hat p^{\phantom{\dagger}}_{j+j_3,2})\Big)\!\Bigg]\,,\\
    &\hat H^{dp}_{12}=\sum_j\Bigg[\,\tilde t^{dp}_{xy}\,\Big(\hat d^\dagger_{j,1}\hat p^{\phantom{\dagger}}_{j,2}+e^{i2\pi/3}\hat d^\dagger_{j+j_2,1} \hat p^{\phantom{\dagger}}_{j,2}\nonumber\\&\hspace{2.5cm}+e^{-i2\pi/3}\,\hat d^\dagger_{j+j_1+j_2,1} \hat p^{\phantom{\dagger}}_{j,2}\Big)\nonumber\\&+\tilde t^{dp}_{xyz}\,\Big(\hat d^\dagger_{j,1} \hat p^{\phantom{\dagger}}_{j+j_3,2}+e^{i2\pi/3}\hat d^\dagger_{j+j_2,1} \hat p^{\phantom{\dagger}}_{j+j_3,2}\nonumber\\&\hspace{2.5cm}+e^{-i2\pi/3}\hat d^\dagger_{j+j_1+j_2,1} \hat p^{\phantom{\dagger}}_{j+j_3,2}\Big)\Bigg]\,,
\end{align}
}
and $\hat H^{dp}_{21}=(\hat H^{dp}_{12})^\dagger$.

\section{Flat-bands within the two-band model}
\label{app:flatband}

\begin{figure}[b!] 
\includegraphics[width = 0.47 \textwidth]{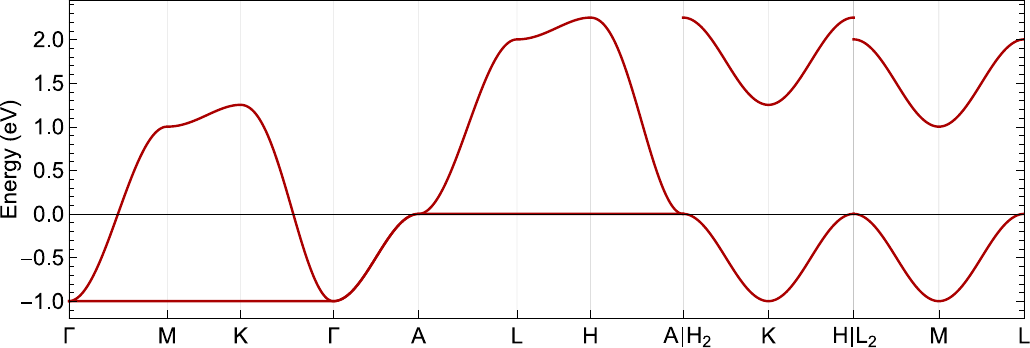}
\caption{
Band structure fulfilling the flat-band conditions.
} \label{fig:FlatBandModel}
\end{figure} 

The eigenvector of the lower band of a generic two-band model reads 
\begin{align}
    \uvec_-(\bk)=\frac{1}{\alpha_\bk}\begin{pmatrix} d_z(\bk)-d(\bk)\\ d_x(\bk)+i d_y(\bk)\end{pmatrix} \, . \label{eqn:v2banddown}
\end{align}
with normalization $\alpha_\bk=\sqrt{2}\sqrt{d(\bk)^2-d_z(\bk)d(\bk)}$, which leads to Eq.~\eqref{eqn:vflat} after inserting the flat-band condition. Considering the Hamiltonian defined in Eq.~\eqref{eqn:HTwoBand}, we construct 
\begin{align}
    d_0(\bk)=\frac{1}{2}\big(H_{11}(\bk)+H_{22}(\bk)\big)\,,\\
    d_x(\bk)=\frac{1}{2}\big(H_{12}(\bk)+H_{21}(\bk)\big)\,,\\
    d_y(\bk)=\frac{i}{2}\big(H_{12}(\bk)-H_{21}(\bk)\big)\,,\\
    d_z(\bk)=\frac{1}{2}\big(H_{11}(\bk)-H_{22}(\bk)\big)\,.
\end{align}
We are interested in a flat band in the $x$-$y$ plane and fix $k_z=0$. Following the procedure as described in Sec.~\ref{sec:flatbandcondition}, we obtain four sets of flat-band conditions different in the relative sign only,
{\allowdisplaybreaks
\begin{align}
    &t^d_{xy+}\!=\!-\frac{1}{2}|\tilde t^d_{xy}|,\, t^d_{xy-}\!=\!-\frac{\sqrt{3}}{2}|\tilde t^d_{xy}|,\, c_0\!=\!\frac{3}{2}|\tilde t^d_{xy}|\!-\!t^d_{z+}\!+\!\mu_d\,, \label{eqn:flatbandconditionUsed}\\
    &t^d_{xy+}\!=\!-\frac{1}{2}|\tilde t^d_{xy}|,\,  t^d_{xy-}\!=\!\frac{\sqrt{3}}{2} |\tilde t^d_{xy}|,\, c_0\!=\!\frac{3}{2}|\tilde t^d_{xy}|\!-\!t^d_{z+}\!+\!\mu_d\,,\\
    &t^d_{xy+}\!=\!\frac{1}{2}|\tilde t^d_{xy}|,\,  t^d_{xy-}\!=\!-\frac{\sqrt{3}}{2} |\tilde t^d_{xy}|,\, c_0\!=\!-\frac{3}{2}|\tilde t^d_{xy}|\!-\!t^d_{z+}\!+\!\mu_d\,,\\
    &t^d_{xy+}\!=\!\frac{1}{2}|\tilde t^d_{xy}|,\, t^d_{xy-}\!=\!\frac{\sqrt{3}}{2} |\tilde t^d_{xy}|,\,c_0\!=\!-\frac{3}{2}|\tilde t^d_{xy}|\!-\!t^d_{z+}\!+\!\mu_d\,.
\end{align}
}
\!\!The dispersion in the $x$-$y$ plane remains flat for any $k_z$ for arbitrary $t^d_{z+}$. A finite $t^d_{z-}$ gaps the quadratic band touching at $k_x=k_y=0$ for any $k_z\neq 0,\pi$ and breaks the flat-band conditions leading to a dispersive lower band in the $x$-$y$ plane. 
Using Eq.~\eqref{eqn:flatbandconditionUsed}, the parameters discussed in Sec.~\ref{sec:CLS} are chosen such that the flat lower band has energy \mbox{$E_-(\bk)=-1$} for $k_z=0$ and $E_-(\bk)=0$ for $k_z=\pi$ with band gap $E_+(\bk)-E_-(\bk)=2$ at momentum $M=(\pi,-\pi/\sqrt{3},0)$.
The Hamiltonian for $t^d_{z-}=0$ explicitly reads 
{\allowdisplaybreaks
\begin{align}
    &d_0(\bk)\!=\!\frac{1}{4}\Big(1\!-\!\cos\!\big((\vb{a}_1\!+\!\vb{a}_2)\!\cdot\!\bk\big)\!-\!\cos\!\big(\vb{a}_1\!\cdot\!\bk\big)\nonumber \\&\hspace{2cm}-\!\cos\!\big(\vb{a}_2\!\cdot\!\bk\big)\!-\!2\cos\!\big(\vb{a}_3\!\cdot\!\bk\big)\!\Big)\, ,\\
    &d_x(\bk)\!=\!\frac{1}{4}\Big(\!\cos\!\big((\vb{a}_1\!+\!\vb{a}_2)\!\cdot\!\bk\big)\!-\!2\cos\!\big(\vb{a}_1\!\cdot\!\bk\big)\!+\!\cos\!\big(\vb{a}_2\!\cdot\!\bk\big)\!\Big)\, ,\\
    &d_y(\bk)\!=\!\frac{\sqrt{3}}{4}\!\Big(\cos\!\big((\vb{a}_1\!+\!\vb{a}_2)\!\cdot\!\bk\big)\!-\!\cos\!\big(\vb{a}_2\!\cdot\!\bk\big)\!\Big)\, ,\\
    &d_z(\bk)\!=\!\frac{\sqrt{3}}{4}\!\Big(\sin\!\big((\vb{a}_1\!+\!\vb{a}_2)\!\cdot\!\bk\big)\!-\!\sin\!\big(\vb{a}_1\!\cdot\!\bk\big)\!-\!\sin\!\big(\vb{a}_2\!\cdot\!\bk\big)\!\Big)\, .
\end{align}
}
\!\!The dispersion is shown in Fig.~\ref{fig:FlatBandModel}.
The normalization of the CLS in Eq.~\eqref{eqn:CLSModel} is
\begin{align}
    &\alpha_\bk^2=2\,\bigg(\!\!-\!3\!+\!\cos\!\big((\vb{a}_1\!+\!\vb{a}_2)\!\cdot\!\bk\big)\!+\!\cos\!\big(\vb{a}_1\!\cdot\!\bk\big)\!+\!\cos\!\big(\vb{a}_2\!\cdot\!\bk\big)\!\bigg)\nonumber\\
    &\times\!\!\bigg(\!\!-\!3\!+\!\cos\!\big((\vb{a}_1\!+\!\vb{a}_2)\!\cdot\!\bk\big)\!+\!\cos\!\big(\vb{a}_1\!\cdot\!\bk\big)\!+\!\cos\!\big(\vb{a}_2\!\cdot\!\bk\big)\nonumber\\
    &\,\,\,\,\,-\!\sqrt{3}\Big(\!\sin\!\big(\vb{a}_1\!\cdot\!\bk\big)\!+\!\sin\!\big(\vb{a}_2\!\cdot\!\bk\big)\!-\!\sin\!\big((\vb{a}_1\!+\!\vb{a}_2)\!\cdot\!\bk\big)\!\Big)\!\bigg) \, ,
\end{align}
which vanishes at $\Gamma=(0,0,0)$. 
\vspace{1.3cm}

\bibliography{main}

\end{document}